# Ultralong Room-Temperature Qubit Lifetimes of Covalent Organic Frameworks

Zhecheng Sun[1,2], Weibin Ni[2,3], Denan Li[3], Xiya Du[1,2], Shi Liu[1,2,3], Lei Sun[1,2,3,*]

[1]Department of Chemistry, School of Science and Research Center for Industries of the Future, Westlake University, Hangzhou 310030, Zhejiang Province, China

[2]Institute of Natural Sciences, Westlake Institute for Advanced Study, Hangzhou 310024, Zhejiang Province, China

[3]Department of Physics, School of Science and Research Center for Industries of the Future, Westlake University, Hangzhou 310030, Zhejiang Province, China

*e-mail: sunlei@westlake.edu.cn

**Abstract**:

Molecular electron spin qubits offer atomic-level tunability and room-temperature quantum coherence. Their integration into engineered solid-state matrices can enhance performance towards ambient quantum information technologies. Herein, we demonstrate covalent organic frameworks (COFs) as programmable matrices of stable organic radical qubits allowing strategic optimization of spin-phonon and spin-spin interactions. Using two classic boronate-ester frameworks, COF-5 and COF-108, to host semiquinone-like radical qubits, we achieve ultralong spin relaxation time ($T_1 >$ 300 μs) at 298 K, which outperforms most molecular qubits and rivals inorganic spin defects. The suppression of spin relaxation is attributed to rigid and neutral structures as well as carbon-centered spin distributions that effectively weaken spin-phonon coupling. Employing dynamical decoupling methods to both COFs improves their quantum coherence and enables room-temperature detection of nuclear spins including $^1$H, $^{11}$B, and $^{13}$C. Our work establishes COFs as designer quantum materials, opening new avenues for quantum sensing of nuclear spins at room temperature.

**Introduction**

Stable organic radicals, e.g. nitroxide,[1,2] semiquinone,[3,4] triphenylmethyl radicals,[5,6] are molecular electron spin qubits with room temperature quantum coherence. They are promising candidates for ambient quantum information technologies and have enabled prototypical demonstrations of molecular quantum logic gates,[7] quantum sensors,[8] and quantum memories.[9] Incorporating stable organic radicals into metal−organic frameworks (MOFs) and covalent organic frameworks (COFs) facilitates rational optimization of their key qubit metrics including spin relaxation time ($T_1$) and decoherence time ($T_2$).[10,11] The corresponding quantum materials, molecular qubit frameworks (MQFs), possess atomically designable, highly ordered, and microporous structures.[12–14] They impart well-defined and fine-tunable spin distributions and phonon dispersion relations, enabling sophisticated control of spin-phonon coupling and spin-spin coupling.[11,15] Their nanoscale pores and functionalizable inner surfaces further facilitate $T_1$ and $T_2$ modulations by harnessing host-guest interactions.[16,17] Recent development of MQFs have led to materials with decent room-temperature spin dynamic properties ($T_1$ = 214 μs, $T_2$ = 0.98 μs),[18] optically initializable quantum states,[19] and quantum sensing capabilities.[20]

Previously, we investigated two MQFs consisting of HOTP-based semiquinone-like radical qubits (HOTP represents 2,3,6,7,10,11-hexaoxytriphenylene) and diamagnetic metal ions, i.e. $Mg_9(HOTP)_4(H_2O)_{30}$ (MgHOTP) and $[CH_3)_2NH_2]_2Ti(HOTP)$ (TiHOTP).[11,16,20] The spin relaxation in these frameworks is closely related to their structural rigidity and innate hydrogen bonds. The former determines density of states (DOS) and cutoff frequencies (Debye frequency, $\omega_D$) of acoustic phonons that drive direct and Raman relaxation processes; the latter affects vibrational modes involving oxygen atoms of HOTP ligands, which stimulate local-mode relaxation processes. Specifically, MgHOTP consists of flexible hydrogen-bonded networks that introduce sub-terahertz optical phonons and expedite spin relaxation. In contrast, the more rigid TiHOTP displays much slower spin relaxation with $T_1$ being 41 μs at 294 K. Soaking it in pyridine further improves the structural rigidity and weakens local hydrogen bonds between HOTP and $(CH_3)_2NH_2^+$, enhancing the room-temperature $T_1$ to 117 μs.

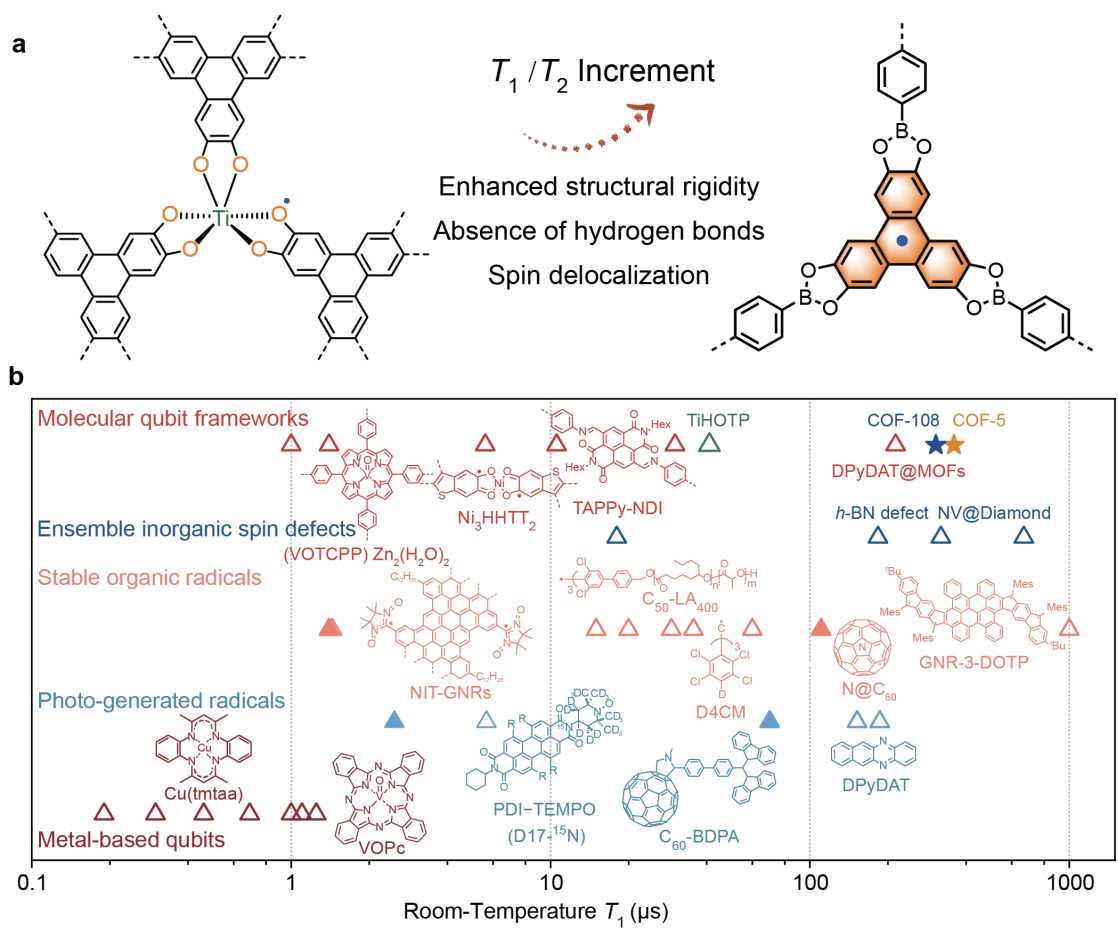

**Figure 1. COFs as solid-state hosts for stable organic radical qubits. a** Advantages of boronate-ester COFs over TiHOTP as MQFs. **b** Comparison of room-temperature $T_1$ values of COF-5 and COF-108 with those of representative molecular electron spin qubits and ensemble paramagnetic defects in dense inorganic solids acquired by pulse EPR spectroscopy. Schemes of selected molecules are shown. Full data and corresponding references are summarized in Supplementary Table 1. Open triangles represent electron spin qubits embedded in solid-state materials or dispersed in solid-state diamagnetic matrices. Filled triangles represent molecular qubits dissolved in fluid solvents. Filled stars represent COF-5 and COF-108.

Further improving $T_1$ of MQFs necessitates implementing stable organic radicals into more rigid structural motifs. This led us to investigate COF-based MQFs as covalent bonds (B−O, C=N, etc.) are stronger than coordination bonds and covalent building blocks are typically less flexible than coordination spheres of metal ions.[21,22] Using π-conjugated monomers could promote spin delocalization, which might help protect spins from environmental vibrational and magnetic influences.[23] Moreover, the frameworks can be neutral, leaving no counterions in the pores, and they can be evacuated to remove adsorbed guest molecules.[24] Thus, COFs could provide rigid matrices without innate hydrogen bonds, which combined would improve $T_1$ and $T_2$ at room temperature (Fig. 1a). In addition, because their structures can be engineered by rational design of monomers and their spin density can be fine-tuned by post-synthetic modification,[10] COFs enable systematic examination about the influence of structural features, e.g., dimensionality, topology, pore size, etc., and spatial distribution of spins on the qubit performance.[25] Overall, COFs could be excellent solid-state hosts for stable organic radical qubits, yet their potential remains largely unexplored, with only one previous example reported to date exhibiting $T_1$ and $T_2$ up to 30.2 μs and 0.49 μs at 296 K, respectively.[10]

Herein, we report ultralong room-temperature $T_1$ of HOTP-based qubits embedded in two classic COFs developed by Yaghi et al., namely COF-5 and COF-108.[26,27] These frameworks exhibit $T_1$ of 359 μs and 306 μs at 298 K, respectively. They outperform most molecular electron spin qubits at room temperature and are comparable with ensemble paramagnetic defects in dense inorganic solids (Fig. 1b). The slow spin relaxation stems from rigid structures, absence of hydrogen bonds, and carbon-centered spin delocalization as revealed by spin dynamic analysis, specific heat capacity measurements, and density functional theory (DFT) calculations. The long $T_1$ leads to $T_2$ reaching 1.3 μs at 298 K, the record for MQFs, and enables room-temperature nuclear spin detection with quantum sensing protocols, showing the potential of radical-integrated COFs for ambient quantum information technologies.[28]

## Results

**Structures and phononic characteristics**

We synthesized COF-5 and COF-108 by reacting methylboronic-acid-protected 2,3,6,7,10,11-hexahydroxytriphenylene (HHTP) with pinacol-protected benzene-1,4-diboronic acid (BDBA) or tetra(4-hydroxyborylphenyl) methane (TBPM) based on literature procedures.[27,29] The reversible transesterification reaction between boronic esters helps reduce density of defects in the products thereby improving their crystallinity. Powder X-ray diffraction (PXRD) confirmed structures of target COFs (Supplementary Fig. 1). COF-5 exhibits a 2D layered structure with honeycomb-like planar sheets and eclipsed interlayer stacking (Fig. 2a).[26] HOTP moieties form a well-ordered lamellar architecture facilitated by π–π interactions. In contrast, COF-108 displays a 3D non-interpenetrated cubic structure nested by triangular HOTP and tetrahedral tetraphenylmethane moieties (Fig. 2a).[27] The π–π interaction is absent between HOTP moieties as they are spatially separated.

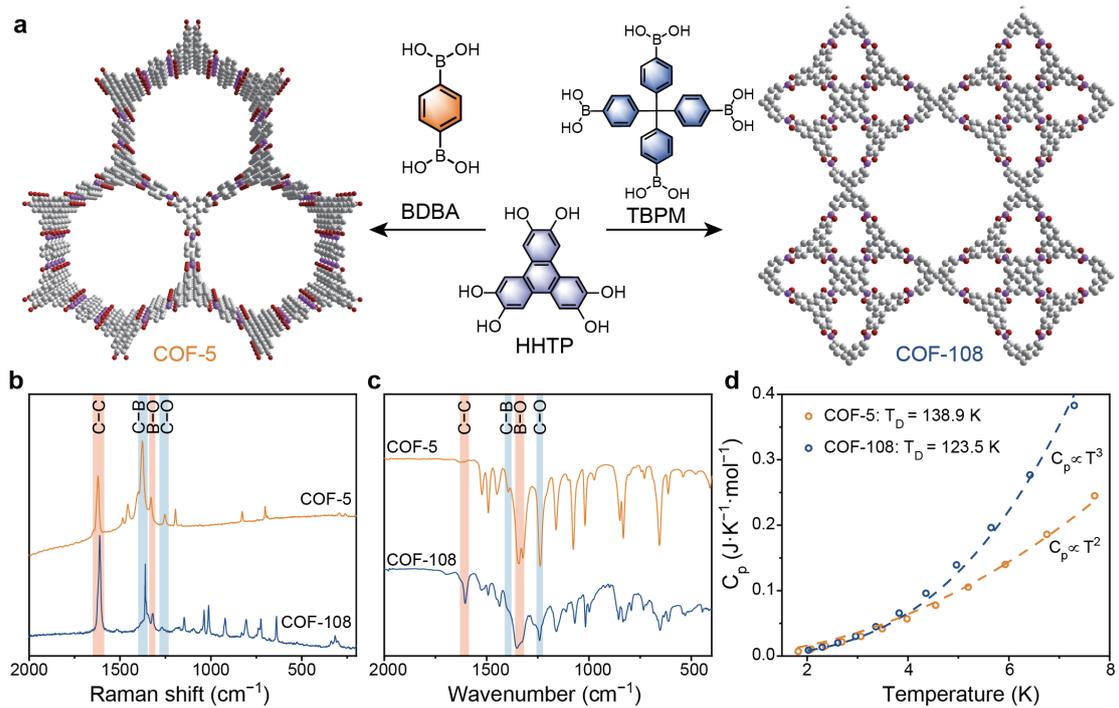

**Figure 2. Crystal structures and phonon characteristics. a** Portions of crystal structures of COF-5 and COF-108. Grey, purple, and red spheres represent C, B, and O, respectively. H atoms are omitted for clarity. **b** Raman spectra of COF-5 and COF-108 collected at 298 K with 785 nm and 633 nm excitation, respectively. Stretch modes of representative chemical bonds are highlighted. **c** FT-IR spectra collected at 298 K. **d** Temperature dependencies of specific heat capacity. Circles are experimental data and dash lines are fitting curves.

Phonon dispersion relations of COF-5 and COF-108 determine their spin-phonon coupling and in turn spin relaxation processes.[30] Specifically, some optical phonons manifest themselves as vibrations of chemical bonds involving spin-bearing atoms, inducing local-mode relaxation whose rate is determined by vibrational frequencies. Acoustic phonons could cause direct relaxation by matching electron Zeeman splitting and could participate in Raman relaxation by providing virtual phononic states. The rates of these two processes decrease with the rise of Debye temperature ($T_D = \hbar\omega_D/k_B$ where $\hbar$ represents reduced Planck constant and $k_B$ represents Boltzmann constant).

Sub-micrometer crystal sizes of COF-5 and COF-108 prevented characterization of their complete phonon dispersion relations by inelastic neutron scattering or resonant inelastic X-ray scattering,[27,31] so we conducted Fourier-transform infrared (FT-IR) and Raman spectroscopy to characterize optical phonons at the Gamma point of 1$^{st}$ Brillouin zone. Both COFs exhibit vibrational peaks at approximately 1250 cm⁻¹, 1330 cm⁻¹, 1377 cm⁻¹, and 1610 cm⁻¹ (Fig. 2b,c), which are respectively attributed to stretch modes of C−O, B−O, C−B, and C−C bonds.[26,32,33] These high-frequency optical phonons may contribute to local-mode relaxation near room temperature.

We further measured specific heat capacity ($C_p$) of COF-5 and COF-108 at various temperatures (T) to acquire their $T_D$. Both frameworks exhibit low electron spin density (*vide infra*) and behave as electronic insulators (Supplementary Fig. 2), so their $C_p$ dominantly stems from phonons. Below 8 K, $C_p$ values of COF-5 and COF-108 scale with $T^2$ and $T^3$, respectively, consistent with Debye models of 2D and 3D systems at the low-temperature limit (T << $T_D$):[34–36]

$$2D: C_p = \frac{24\zeta(3)N_A k_B T^2}{T_D^2} \quad \text{eq.1}$$

$$3D: C_p = \frac{12\pi^4 N_A k_B T^3}{5T_D^3} \quad \text{eq.2}$$

where $N_A$ represents Avogadro's number and $\zeta(x)$ is Riemann zeta function. These observations indicate consistent distribution of phonon density of states with structural dimensionality for each COF. Fitting the temperature dependencies of $C_p$ with Equation (1) and (2) revealed $T_D$ values of 138.9 K and 123.5 K for COF-5 and COF-108, respectively (Fig. 2d). As $T_D$ is a proxy for structural rigidity, COF-5 is slightly more rigid than COF-108 possibly due to the interlayer π−π stacking in the former.

**Room-temperature qubit performance**

X-band (9.8 GHz) continuous-wave electron paramagnetic resonance (CW-EPR) spectroscopy revealed slight *g*-anisotropy for each framework with COF-5 showing $g_{\parallel}$ = 2.00313 and $g_{\perp}$ = 2.00351 while COF-108 showing $g_{\parallel}$ = 2.0031 and $g_{\perp}$ = 2.00353 (Supplementary Fig. 3a,b). These values are comparable with the free-electron *g*-factor ($g_e$ = 2.0023), confirming the radical characteristics of electron spins. The radical concentrations amount to $5.2 \times 10^{14}$ spins·mm$^{-3}$ (0.095% HOTP) and $9.3 \times 10^{14}$ spins·mm$^{-3}$ (0.117% HOTP) in COF-5 and COF-108, respectively, as indicated by quantitative CW-EPR measurements. Both radical percentages are lower than that of HHTP used for the COF synthesis (0.202% HHTP) (Supplementary Fig. 3c), demonstrating that the transesterification reaction reduces the density of radical defects, i.e. partially oxidized HOTP moieties. DFT calculations show that the electron spin mainly resides on triphenylene moieties in both COFs (Fig. 3e,f), exhibiting distinct spin distribution from that of the HOTP radical in MgHOTP and TiHOTP where electron spin is concentrated on oxygen atoms.[11] This difference is likely caused by the electron donation of HOTP to electron-deficient boron, which significantly reduces the electron density on oxygen atoms and leaves the unpaired electron mainly distributed on the aromatic rings. In addition, DFT calculations indicate efficient spin delocalization along the π-stacked HOTP columns in COF-5 (Supplementary Fig. 4), contrasting with localized spin on single HOTP moiety in COF-108. Thus, comparing these two COFs could reveal the influence of structural dimensionality and its associated spin delocalization among HOTP moieties on spin dynamics.

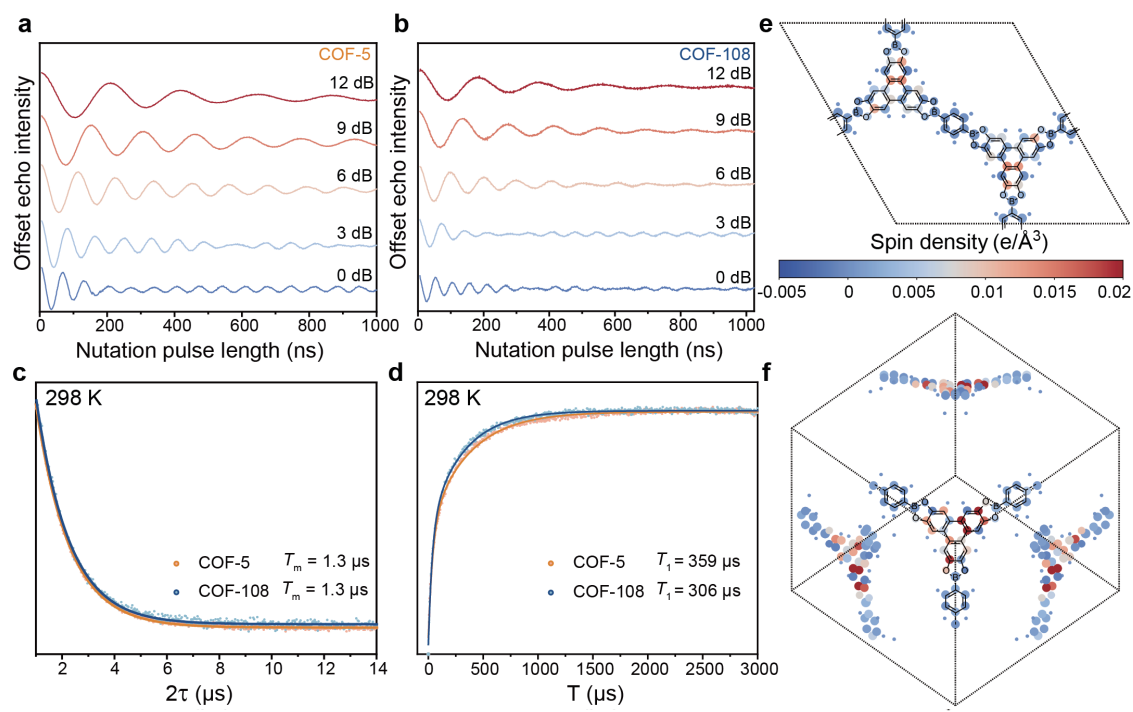

**Figure 3. Room-temperature (298 K) qubit performance. a,b** Rabi oscillations of COF-5 and COF-108. Microwave attenuations are labeled for nutation traces. **c** Hahn echo decay measurements of $T_m$. Dots are experimental data. Solid lines are mono-exponential decay fitting curves. **d** Inversion recovery measurements of $T_1$. Solid lines are bi-exponential decay fitting curves. **e,f** Spin density distributions of COF-5 and COF-108.

We employed X-band pulse EPR spectroscopy to characterize $T_1$ and phase memory time ($T_m$), the latter of which encompasses all dephasing factors and represents $T_2$ for ensemble systems, of COF-5 and COF-108 at room temperature (298 K). Both COFs exhibited echo-detected field sweep spectra corresponding to $S = 1/2$ spins (Supplementary Fig. 5) as well as Rabi oscillations in nutation experiments (Fig. 3a, 3b, and Supplementary Fig. 6), demonstrating their coherent addressability and qualifying them as MQFs. Their $T_m$ times are approximately 1.3 μs as revealed by Hahn echo decay measurements (Fig. 3c), which are longer than room-temperature $T_m$ of other MQFs (Supplementary Table 1). Inversion recovery measurements showed comparable behaviors of the two COFs involving two exponential decays. The faster is likely caused by spectral diffusion, whereas the slower is attributed to intrinsic spin relaxation and its characteristic decay time is assigned to $T_1$. Remarkably, $T_1$ times of COF-5 and COF-108 are 359 μs and 306 μs, respectively (Fig. 3d). They are significantly longer than the room-temperature value of TiHOTP ($T_1$ = 41 μs). To our knowledge, these two COFs show comparable $T_1$ at room temperature with ensemble paramagnetic defects in inorganic solids (e.g. nitrogen−vacancy centers in diamond[37]) and three carbon-centered radicaloids dispersed in *o*-terphenyl,[38] and they outperform all other molecular electron spin qubits encompassing stable organic radicals,[28,39] photo-generated radicals,[40] and coordination complexes[41] (Fig. 1b and Supplementary Table 1).

**Spin relaxation mechanisms**

To articulate the spin relaxation mechanisms, we measured $T_1$ of COF-5 and COF-108 in the temperature range of 30 K – 298 K. These two materials exhibited almost identical $T_1$ at each temperature (Fig. 4a), indicating that the structural dimensionality and in turn the spin delocalization among HOTP moieties do not affect spin relaxation. $T_1$ increases monotonically with decreasing temperature in each COF: it increases from 300 – 360 μs at 298 K to 130 – 170 ms at 30 K. It cannot be accurately measured below 30 K due to instrumental limitations, yet it has the potential to exceed 1 s below 10 K (see Supplementary Table 2 and 3). The spin relaxation rate ($1/T_1$) scales with $T^{2.9}$ (Supplementary Fig. 7), indicating the presence of multiple two-phonon processes driven by both acoustic and optical phonons.[42] As the electron spin density mainly distributes on the triphenylene moieties, C−C stretches should make a significant contribution to the spin relaxation. In addition, considering the low spin density and relatively high experimental temperatures, both cross relaxation and direct process should be negligible.[11]

Based on this analysis, we fitted the temperature dependence of $1/T_1$ with a Raman process driven by acoustic phonons and a local-mode process driven by C−C stretches:

$$\frac{1}{T_1} = A_{Ram}\left(\frac{T}{T_D}\right)^9 \int_0^{T_D/T} \frac{x^8 e^x}{(e^x-1)^2}dx + A_{Loc}\frac{e^{h\nu/k_BT}}{(e^{h\nu/k_BT}-1)^2} \ldots\ldots\ldots\ldots\text{eq.3}$$

where $A_{Ram}$ and $A_{Loc}$ are pre-factors of Raman and local-mode processes, respectively,

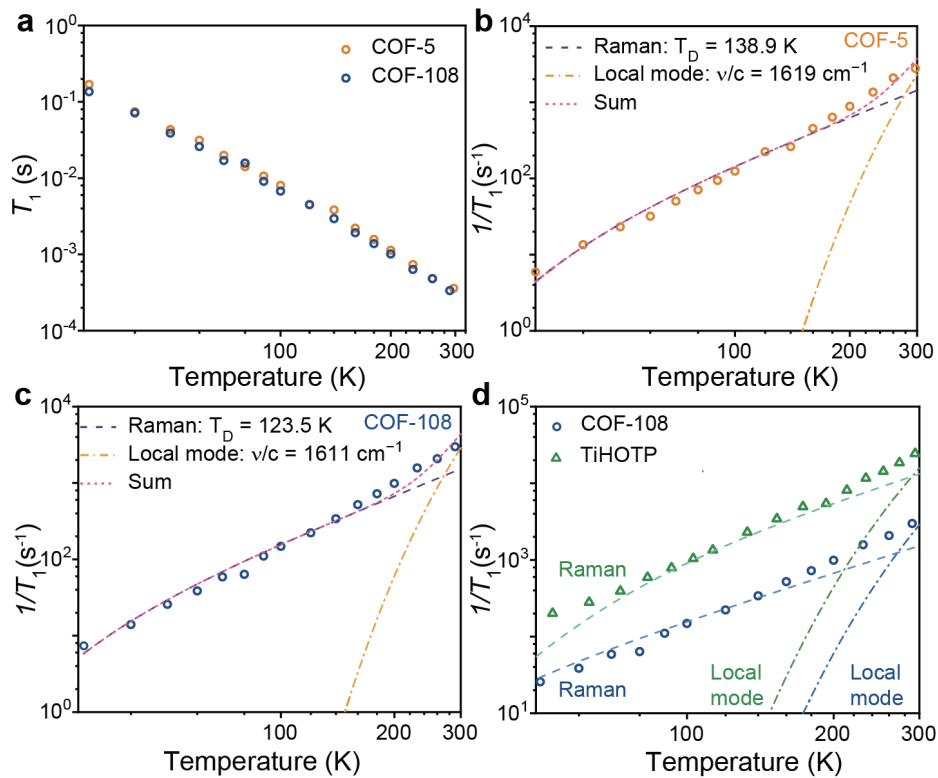

**Figure 4. Spin relaxation mechanisms. a** $T_1$ of COF-5 and COF-108 acquired at various temperatures. **b,c** Spin relaxation rates ($1/T_1$) acquired at various temperatures and their fitting results based on Equation 3. Circles represent experimental data. Dash lines represent contributions from individual spin relaxation processes. Debye temperatures and vibrational frequencies used for fitting are indicated. **d** Comparison between spin relaxation behaviors of COF-108 and TiHOTP. Contribution of a direct process was omitted for TiHOTP for clarity. Data of TiHOTP were extracted from Ref. 11.

and ν represents the linear frequency of optical phonon. $T_D$ and ν were extracted from specific heat capacity and vibrational spectroscopic measurements (*vide supra*), respectively, reducing fitting parameters to the two pre-factors. This equation yielded decent fitting results (Fig. 4b,c). For both COFs, the Raman process is dominant below 250 K with $A_{Ram}$ being approximately $2 \times 10^3$ s$^{-1}$, and the local-mode process dominates above this temperature with $A_{Loc}$ being approximately $6 \times 10^6$ s$^{-1}$. Notably, both pre-factors are smaller than those observed for TiHOTP and many semiquinone radicals[11], making both relaxation processes in COFs significantly slower (Fig. 4d). The suppression of Raman process is likely due to enhanced structural rigidity because B−O bonds and [BO$_2$C] linkers are more rigid than Ti−O bonds and [TiO$_6$] coordination spheres, respectively. The local-mode process may be inhibited for two reasons. First, the formation of boronic ester shifts spin distribution in HOTP from peripheral oxygen atoms to its triphenylene core, making spins less prone to the influence from guest molecules. This also leads to spin delocalization on triphenylene that reduces spin density on each atom, so spins are less affected by local vibrations of chemical bonds. Second, both COF-5 and COF-108 are neutral. Evacuating them during sample preparation could greatly reduce the amount of guest molecules adsorbed in the pores and in turn eliminate local hydrogen bonds involving HOTP. These combined effectively suppress spin relaxation in both COFs, giving rise to the exceptionally long $T_1$ at room temperature.

**Decoherence in COFs and quantum sensing capabilities**

We further probed spin decoherence mechanisms of COF-5 and COF-108 by conducting Hahn echo decay measurements from 30 K to 298 K (Fig. 5a). The $T_m$ values of these two materials are comparable across this temperature range. They remain almost constant below 120 K, indicating that the decoherence is governed by temperature-independent nuclear spectral diffusion, electron spin flip-flop, and/or instantaneous diffusion.[43,44] Above this temperature, $T_m$ drops acutely with the rise of temperature. Mechanistic analysis revealed the dominant role of electronic spectral diffusion above 130 K for both COFs (Supplementary Fig. 8): spin relaxation of nearby electron spins perturbs local magnetic environment and accelerates decoherence.[45] Thus, $T_1$ poses the upper limit for $T_m$ above 120 K even though the former is much longer. The decent $T_m$ values observed at room temperature for both COFs benefit from their exceptionally long $T_1$. In addition, because the contribution of electronic spectral diffusion decreases as spin-spin coupling declines, reducing spin density would further improve $T_m$ of COFs.[10,43]

We employed a classical dynamical decoupling method, Carr−Purcell−Meiboom−Gill (CPMG) pulse sequence, to extend $T_m$ at room temperature.[46,47] The CPMG sequence consists of a series of spin-locking π pulses applied after an initial π/2 excitation pulse, which refocus the spins and allow for the measurement of spin dynamics over time. This refocusing helps mitigate the effects of inhomogeneous magnetic fields, effectively suppressing the decoherence caused by electronic and nuclear spectral diffusions. Take COF-108 as an example. Compared with the room-temperature value observed by the Hahn echo decay sequence ($T_m$ = 1.3 μs), applying a CPMG sequence consisting of two π pulses (CPMG-2) improves $T_m$ to 1.6 μs, employing eight π pulses (CPMG-8) extends $T_m$ to 5.5 μs, and using more π pulses does not lead to further improvement (Fig. 5b and Supplementary Fig. 9a).

Applying CPMG-8 to COF-5 improved its Tm from 1.3 μs to 2.9 μs. (Supplementary Fig. 9b). Hence, $T_m$ of COFs can be improved by dynamically suppressing environmental magnetic noise.

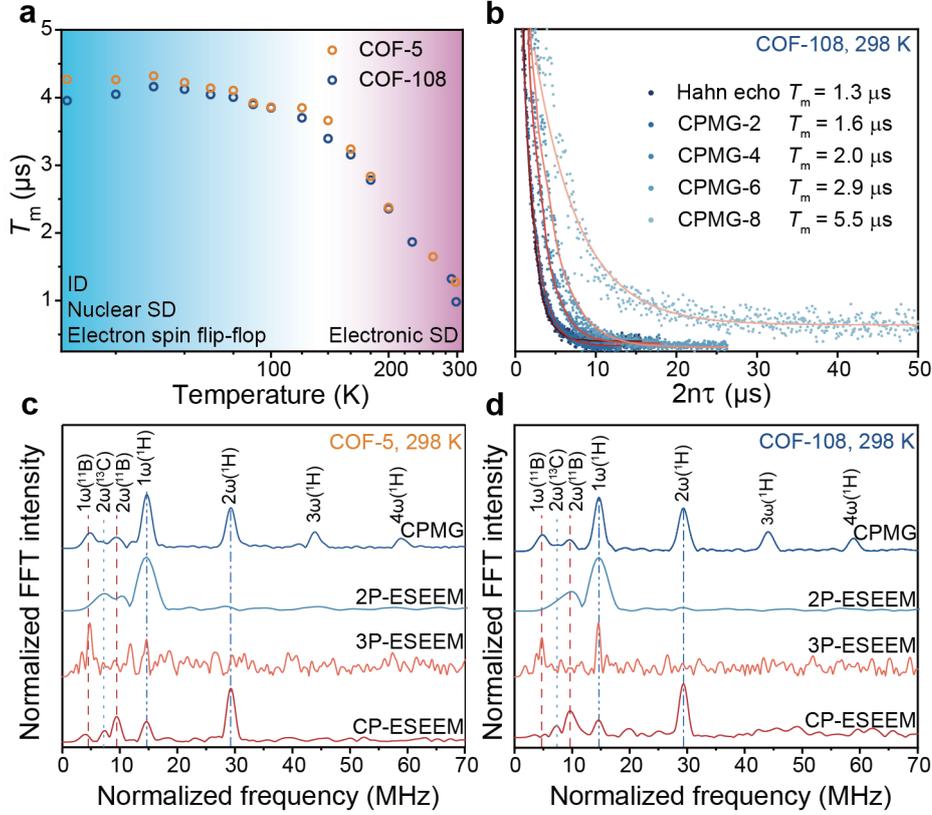

**Figure 5. Spin decoherence and nuclear spin modulation. a.** $T_m$ of COF-5 and COF-108 acquired at various temperatures. SD represents spectral diffusion; ID represents instantaneous diffusion. **b** Hahn echo decay and CPMG measurements of $T_m$ at 298 K for COF-108. Circles represent experimental data. Solid lines are mono-exponential decay fitting curves. **c,d** Frequency-domain spectra acquired by CPMG-16 and various types of ESEEM sequences at room temperature for COF-5 and COF-108. The $n^{th}$ harmonics of the Larmor frequency of a nuclear spin X is represented as nω(X). Frequencies were normalized to 1ω($^1$H).

In addition to prolonging $T_m$, the CPMG sequence can also detect alternative-current (AC) magnetic fields. Indeed, employing it to COF-5 and COF-108 generated decay curves with pronounced spectral oscillations (Fig. 5b). These oscillations are manifestations of the electron spin echo envelope modulation (ESEEM).[48] They originate from the modulation of electron spin precession in COFs by nuclear spin precession. Under the weak coupling regime where the hyperfine constant is much smaller than the nuclear Larmor frequency, the modulation frequency reflects gyromagnetic ratio of surrounding nuclear spins, and the modulation depth corresponds to their quantities. As the gyromagnetic ratio is an intrinsic property of a nucleus and it differs for different elements and isotopes, it unambiguously reveals the identity of nucleus. Thus, ESEEM signals can be used to identify and quantify nuclear spins around the electron spin, potentially enabling quantum sensing applications of COF-5 and COF-108.[20]

To further investigate the quantum sensing capability of COFs as well as to articulate environmental nuclear spins that act as decoherence sources, we compared the nuclear spin modulation signals at room temperature obtained from Hahn echo decay (also called 2-pulse (2P) ESEEM), 3-pulse (3P) ESEEM, combination-peak (CP)-ESEEM,[20] and CPMG-16 sequences (Supplementary Fig. 10).[49] The long $T_1$ and $T_m$ of COF-5 and COF-108 enable the application of these quantum sensing sequences at room temperature. Both COFs exhibit clear modulation features from $^1$H, $^{11}$B, and $^{13}$C nuclei under these sequences, demonstrating these nuclei as sources of nuclear spectral diffusion (Fig. 5c,d and Supplementary Fig. 8). 2-pulse ESEEM generates broad peaks with unresolved low-frequency features, and 3-pulse ESEEM shows weak modulations with low signal-to-noise ratio (SNR). CPMG-16 gives rise to 1st to 4th harmonics of $^1$H, but modulations from $^{11}$B and $^{13}$C appear as a combined peak, the latter of which are clearly resolved with CP-ESEEM (Fig. 5c,d). The observation of $^{11}$B at its nuclear Larmor frequency (or 2nd harmonics) indicates negligible hyperfine interaction between HOTP radicals and $^{11}$B nuclear spins.[48] This is consistent with the results of DFT calculations: as the electron spin primarily distributes on the conjugated triphenylene moiety, it should interact with $^{11}$B nuclei through dipolar coupling rather than Fermi contact, resulting in weak hyperfine interaction. Overall, these results confirm the capability of COFs for nuclear spin detection and indicate their potentials for quantum sensing of nuclear spin, which might empower unambiguous identification and precise quantification of guest molecules.

Comparing the four quantum sensing sequences, we find that increasing the number of π pulses effectively refocuses electron spins, counteracts phase diffusion, and enhances and accumulates periodic modulation signals, thereby significantly improving the SNR for sensing. For the same sample, the CPMG sequence exhibits the strongest modulation depth and the best SNR but has limited capability to resolve low-frequency signals due to insufficient $T_m$. In contrast, the CP-ESEEM sequence is effective at resolving low-frequency signals by harnessing the long $T_1$ but has limited SNR as a result of its weaker modulation depth. Developing new pulse sequences that integrate the benefits of CPMG and CP-ESEEM holds promise for further improving the sensitivity, spectral resolution, and dynamic range for quantum sensing applications.

**Conclusion**

In conclusion, COF-5 and COF-108 exhibit excellent qubit performance at room temperature. The rigid structures, triphenylene-centered spin distributions, and absence of hydrogen bonds in these two frameworks together suppress spin relaxation. This leads to $T_1$ above 300 μs at 298 K that is longer than the values observed for most molecular electron spin qubits. The ultralong $T_1$ gives rise to $T_m$ reaching 1.3 μs at room temperature, setting the record for MQFs. The excellent qubit performance enables applications of dynamical decoupling and ESEEM sequences, showing the capability of COF-5 and COF-108 for nuclear spin detection. The foregoing results demonstrate that COFs provide viable solid-state hosts for molecular electron spin qubits. Integrating design principles unveiled in this study with well-established synthetic methodologies for radical-embedded COFs could lead to MQFs with long $T_1$ and $T_m$ at room temperature.[50–52] Rational design and post-synthetic modification of the inner surfaces of COFs would further generate candidate materials for quantum sensing of nuclear spins in ambient conditions.

**Methods**

**Synthesis**

COF-5 was synthesized based on literature procedures.[29] COF-108 was synthesized with modified literature procedures.[29] Tetrakis(4-(4,4,5,5-tetramethyl-1,3,2-dioxaborolan-2-yl)phenyl)methane (24.72 mg, 0.03 mmol), methylboronic acid (96 mg, 1.6 mmol), HHTP (13.08 mg, 0.04 mmol), 0.5 mL of mesitylene, 0.5 mL dioxane, and 50 μL of trifluoroacetic acid were added to a 10 mL glass ampoule. The reaction was frozen by liquid $N_2$ and then evacuated for 5 min. The ampoule was sealed under vacuum and was heated at 120 °C for 4 days to afford COF-108 as white precipitates. The product was cleaned by anhydrous and deoxygenated tetrahydrofuran with Soxhlet extraction for 24 h. It was dried under vacuum and was kept in a $N_2$-filled glovebox.

**Vibrational spectroscopy**

FT-IR spectroscopy was conducted on a Nicolet iS50 spectrometer (Thermo Fisher

Scientific) equipped with an attenuated total reflectance (ATR) module.

Raman spectra were acquired by WITec Alpha300RAS spectrometer. 785 nm and 633 nm excitation lasers were used for COF-5 and COF-108, respectively, to avoid background fluorescence while retaining good spectral resolution and signal-to-noise ratio.

**Continuous-wave (CW) EPR spectroscopy**

Samples of COF-5 or COF-108 were added into quartz EPR tubes (i.d. = 2 mm) that were flame-sealed under vacuum. They were used for EPR characterization. CW-EPR spectroscopy was performed on a Bruker ELEXSYS-II E500 spectrometer at X-band (9.8 GHz). The spectra were collected at 298 K with a modulation amplitude of 0.05 mT and a microwave power of 0.1 mW. Each spectrum of COFs was fitted by EasySpin 6.0.0 in MATLAB R2023b with one type of isotropic electron spin ($S = 1/2$) and Gaussian lineshape.[53] The spectrum of HHTP was fitted with two types of isotropic electron spins ($S = 1/2$) and Lorentzian lineshape. Quantitative analysis was conducted using X-EPR Quantitative Analysis, SpinCount Module.[10]

**Pulse EPR spectroscopy**

Pulse EPR spectroscopy was performed on a CIQTEK EPR100 spectrometer at X-band (9.6 GHz) with EPR-VTS-L-D1-PWC closed-loop helium cryostat and Lakeshore336 temperature controller. 512 data points and 32 ns π pulses were used without further notice.

*Inversion recovery pulse sequence* ($\pi - T - \pi/2 - \tau - \pi - \tau -$ echo) was applied to measure $T_1$ with 1024 data points. The length of T started at 400 ns with different step increments that vary with temperature. τ was fixed at 150 ns. Background noise was canceled by four-step phase cycling with pulse phases of (+x, −x, +x) (+x, +x, +x) (−x, −x, +x) and (−x, +x, +x). Inversion recovery curves were fitted by the following biexponential decay function to extract $T_1$:

$$I = A_S e^{-T/T_S} + A e^{-T/T_1} + I_0 \dots\dots\dots\dots\dots\dots\dots\dots\dots\text{eq. 3}$$

where I is echo intensity, $A_S$ and $A$ are pre-factors, $T_s$ describes a fast relaxation process possibly stemming from spectral diffusion, and $I_0$ accounts for the baseline drift.

*Hahn echo decay pulse sequence* ($\pi/2 - \tau - \pi - \tau -$ echo) was applied to measure $T_m$.

The length of τ started at 150 ns with different step increments that vary with temperature. Background noise was canceled by two-step phase cycling with pulse phases of (+x, +x) and (−x, +x). Hahn echo decay curves were fitted by the following exponential decay function to extract $T_m$:

$$I = Ae^{\frac{2\tau}{T_m}} + I_0 \dots\dots\dots\dots\dots\dots\dots\dots\dots\dots\dots\dots\text{eq. 4}$$

*CPMG-n pulse sequence* ($\pi/2_x - (\tau - \pi_y - \tau)_n -$ echo; x and y refer to pulse phases) was applied to measure $T_m$ and to detect nuclear spins. The length of τ started at 150 ns with different step increments that vary with n. For the former purpose, n was chosen as 2, 4, 6, 8. Background noise and unwanted echoes were canceled by complete phase cycling proposed by Lombardi et al.[47] In this method, the phase of each pulse was cycled, allowing the measurement of refocused echo after $2^{n+1}$ phase cycling steps. CPMG decay curves were fitted by the following exponential decay function to extract $T_m$:

$$I = Ae^{\frac{2n\tau}{T_m}} + I_0 \dots\dots\dots\dots\dots\dots\dots\dots\dots\dots\dots\dots\text{eq. 5}$$

For the latter purpose, n was chosen as 16 and 256 data points were collected. τ started at 150 ns and was incremented 4 ns per step. Two-step phase cycling, where only the phase of the first pulse was cycled, was used to cancel background noise. Integration of the resulting echo was plotted against the delay time, giving an oscillatory time-domain ESEEM spectrum. This was background corrected with polynomial fitting, apodized with the Hamming window function, zero-filled, and transformed to frequency domain by FFT in EPR ProC 5.1.4.

**DFT calculations**

DFT calculations were performed using the CP2K software package with the QuickStep module.[54] The generalized gradient approximation of Perdew-Burke-Ernzerhof (PBE) for the exchange-correlation functional was employed,[55] augmented with Grimme's DFT-D3 dispersion correction to account for van der Waals interactions.[56] A double-zeta valence polarized (DZVP) basis set was employed in conjunction with Goedecker-Teter-Hutter (GTH) pseudopotentials.[57,58] The energy cutoff was set to 800 Ry. The self-consistent field (SCF) convergence criterion was $10^{-6}$. During the structural optimizations, the force and pressure convergence criteria were $4.5 \times 10^{-4}$ Hartree/Bohr and 100 bar, respectively. Convergence of the total energy with respect to supercell dimensions and energy cutoff was verified. In the initial configuration, the oxygen atom was assigned a formal charge of +1 to simulate the unpaired electron environment characteristic of the free radical species. Mulliken population analysis was utilized to quantify the atomic contributions to the spin density distribution.[59]

## Data availability

The data supporting our findings are included in the article and supplementary files. Additional data generated during the study are available from the corresponding author upon request.

## Acknowledgments

This work was supported by the National Natural Science Foundation of China (Grant No. 22273078) and the Hangzhou Municipal Innovation Team Program (TD2022004). We are grateful to Dr. Zhongyue Zhang and Aimei Zhou for valuable discussions. Z.S. acknowledges Danyu Gu and Dr. Shutian Lu for assistance with EPR spectroscopy as well as Dr. Shaoze Wang and Dr. Chao Zhang for assistance with specific heat capacity measurements and data analysis. Z.S. and L.S. thank the Instrumentation and Service Center for Molecular Sciences and the Instrumentation and Service Center for Physical Sciences at Westlake University for providing facility access and technical support. Computational resources were provided by the Westlake HPC Center.

## Author contributions

L.S. and Z.S. initiated the project and designed experiments. L.S. provided overall supervision. Z.S. conducted material synthesis, EPR characterization, vibrational spectroscopy, and specific heat capacity measurements. W.N. and X.D. assisted with EPR spectroscopy. D.L. and S.L. carried out DFT calculations. The manuscript was co-written by L.S. and Z.S. All authors reviewed and approved the final version of the manuscript.

## Competing interests

The authors declare no competing interest.

## Additional information

**Supplementary information:** The online version contains supplementary information available at https://

**Correspondence** and requests for materials should be addressed to Lei Sun.